\documentclass[slac_one]{revtex4}
\usepackage{graphicx}
\usepackage{fancyhdr}
\begin{document}

\title{New Methods for the Calculation of Multi-Loop Amplitudes} 

%

\author{P. Marquard}
\affiliation{Institut f\"ur Theoretische Teilchenphysik, Universit\"at
  Karlsruhe, D-76131 Karlsruhe, Germany}

\begin{abstract}
We present a brief review of current methods for the  calculation of
multi-loop amplitudes including recent developments. As an example we
present the calculation of the second moment of the heavy quark current
correlator and the extraction of the values of the charm and bottom
quark masses using the results of this calculation.
\end{abstract}

\maketitle
\thispagestyle{fancy}


\section{INTRODUCTION AND MOTIVATION} 
\label{sec:intr-motiv}
In the era of the LHC and in preparation for the ILC there is need
for multi-loop calculations to match the precision of these experiments
from the theory side. The classes of problems currently accessible reach
from multi-leg multi-scale problems at one-loop up to no-scale
propagator, and tadpole diagrams at four-loop level. Many problems fall
into these categories directly while others can be cast into the
required form by appropriate expansions in small parameters. A typical
multi-loop calculation proceeds as follows: The needed diagrams are
generated and if necessary expanded to decrease the difficulty of the
problem. The resulting Feynman integrals are then reduced to a small set
of simpler integrals, which have to be calculated in a further step.
While the reduction of the integrals can be automatized completely and
is mostly independent of the problem at hand, the calculation of the
remaining integrals must still be done manually for every calculation.
In the following we give a brief review of the current status of the
reduction to master integrals which with recent developments is now in a
good shape such that the calculation of the master integrals now again
poses the most difficulties in multi-loop calculations.

\section{TECHNICALITIES}
\label{sec:technicalities}
One of the main ingredients of many multi-loop calculations are the
so-called Integration-By-Parts (IBP) identities \cite{Chetyrkin:1981qh}
\begin{equation}
  \label{eq:1}
0 =  \int d^nl_1 \cdots d^nl_m \frac{\partial}{\partial l^\mu_j} \frac
  {\{l^\mu_i,p^\mu_i\}}{P_1^{k_1} \cdots P_n^{k_n}} \, ,
\end{equation}
which hold in dimensional regularization. Here we consider typical
Feynman integrals, where $l_i$ are loop and $p_i$ are external momenta.
These identities can be
used to relate different integrals. Using these relations the full set
of needed integrals, which can be of the order of $10^4-10^6$
integrals, can be reduced to a small set of so-called master
integrals. In a typical calculation there appear ${\cal O}(10-100)$
master integrals, which remain to be calculated directly. Another
main advantage of the reduction to master integrals is that many
properties like gauge invariance can be checked on the level of master
integrals, since these are linearly independent.

To make use of the IBP identities the Laporta algorithm \cite{Laporta:2001dd}, a
Gauss-elimination-like algorithm, is frequently used. In order to use
the Laporta algorithm, one has to choose an ordering to ensure that
more difficult integrals are expressed through simpler ones. Once this
ordering has been chosen the algorithm proceeds as follows: Choose a
set of integrals, which you like to solve; pick an integral from the
set and generate an IBP; solve for the most difficult integral in the
equation and continue with the next IBP or integral.  
The system of equations will become overdetermined if one chooses a
sufficiently large set of integrals, since there are
$N_L (N_L + 2 N_E +1)/2$ IBP identities generated by a given integral, where $N_L$ and $N_E$ are the  number of loops
and legs, respectively.

Although this method has been successfully applied in many calculations
it has some drawbacks. The complexity of the Laporta algorithm is ${\cal
  O}(n^3)$, 
where $n$ is the number of equations. In combination with the
combinatorial properties of the problem this leads to fast growth of
the size of the problem to be solved. Furthermore in typical
applications the system of equations is overdetermined by a factor 3-5, 
so quite some time is spent just checking that the system of equations
is consistent.

A different approach for solving the set of IBP identities is provided
by Groebner bases. Groebner bases arose in the context of ideals of
rings over multivariate polynomials. More precisely, they were
introduced while trying to solve the problem, how to decide if a certain
polynomial belongs to an ideal of a ring of polynomials. An ideal $I$ is
a subset of the ring $R$ such that for every $q\in I, r\in R$, the product is
an element of $I$, $q \cdot r \in I$.

An ideal can be generated by a set of polynomials $Q=\{q_1,\ldots,q_n\}$
using $I = \{p \in R | p = \sum r_i q_i , r_i \in R\}$. Given these
generating polynomials, the question whether a given polynomial $p\in R$ is an
element of the ideal, can easily be answered in the case of univariate
polynomials, where the operation of polynomial division is
unambiguous. If the polynomial division by the generating polynomials
leaves no remainder, $p$ is an element of the ideal. 

This statement does  no longer hold, when dealing with multivariate
polynomials. Here the remainder of multivariate divisions by multiple
polynomials  depends on the chosen ordering of the
polynomials. Therefore if the remainder of such divisions is not
zero one may not conclude that the polynomial is not an element of  the
ideal. 

At this point Groebner bases prove very useful since a multivariate
division with respect to elements of a Groebner basis is again
unambiguous. This leads to one of the definitions of a Groebner basis: A set $B
\subset R$ is a Groebner basis if the reduction, i.e. multivariate
division, of any $r\in R$ with respect to the elements of $B$ yields the
same remainder independent of the chosen order. Using the Buchberger
algorithm a Groebner basis can be constructed from any set of polynomials
generating the ideal.

This knowledge can be used for the solution of the IBP-identities. For
this purpose the IBP identities have to be written in terms of shift
operators \cite{Smirnov:2005ky,Smirnov:2006tz,Smirnov:2008iw} or
differential operators \cite{Tarasov:1998nx,Tarasov:2004ks}. The IBP relations then form a left-ideal in the
now non-commutative ring of shift operators for which a Groebner basis
has to be constructed. Every integral can be obtained by applying shift
operators to a basis integral, i.e., $J(a,b) = p J(1,1),p\in R$.  Since
every element $p$ of the ring can be expressed in the form $p = \sum r_i
p_i + q,$ where $p_i\in I , r_i,q\in R$ the integral can be reduced via
$J(a,b) = p J(1,1) = ( \sum r_i p_i + q ) J(1,1) = q J(1,1)$, since $r_i
p_i \in I$ and therefore $r_i p_i J(1,1) = 0$.

Although this procedure does, in principle, generate a reduction of any
integral there are still some problems. The Buchberger algorithm is
guaranteed to generate a basis but it might need a very long time and a
lot of resources. In practice there is no implementation of the
Buchberger algorithm which succeeds in constructing a basis for
non-trivial problems. Furthermore, as can be checked in simple cases, the
application of Groebner bases to the reduction problem does not yield
the minimal number of master integrals that can be reached using a
Laporta-like kind of reduction method.

To circumvent the problematic issues of the Groebner basis approach, a
modification of the Buchberger algorithm has been introduced to
construct so-called s-bases. Unfortunately in this approach the
construction of a basis is no longer guaranteed to succeed and is highly
dependent on the chosen ordering. For certain topologies a basis cannot
be found and therefore the method cannot be applied.  In this cases a
conventional Laporta approach still has to be used in the sectors where
no basis can be found.

\section{EXAMPLE AND APPLICATION}
\label{sec:example-application}
As an example for the power of the methods outlined in the previous
Section we present the calculation of the second moment of the vector
correlator at ${\cal O}(\alpha_s^3)$ and its application to the
determination of the charm and bottom quark masses from experimental
data. 

The charm and bottom masses can be extracted from the $R$-ratio
$R(s)=\frac{\sigma(e^+e^-\to hadrons)}{\sigma(e^+e^- \to \mu^+
  \mu^-)}$ \cite{Shifman:1978by,Reinders:1984sr}. The extraction of the quark masses can be done by
considering the moments $C_n^{{exp}}$ of the ratio defined by 
\begin{equation}
  \label{eq:4}
  C_n^{{exp}} = \int \frac{R(s)}{s^{n+1}} ds \, .
\end{equation}
These moments are related to the Taylor coefficients $C_n^{theo}$ of the
low energy expansion of the heavy quark current correlator
\begin{equation}
  \label{eq:2}
\sum_{n=0}^\infty C_n^{theo} \left ( \frac{q^2}{4 M^2}\right )^n  = \Pi(q^2) \, ,
\end{equation}
where $\Pi$ is the vacuum polarization function of the photon
\begin{equation}
  \label{eq:3}
  \Pi_T (q^2 g^{\mu\nu} - q^\mu q^\nu)  = \Pi^{\mu\nu} = \langle 0| T j^\mu(q) j^\nu(0)| 0 \rangle
\end{equation}
with the heavy quark current $j^\mu = \bar \psi \gamma^\mu \psi$. These
theoretical moments can be calculated in an expansion in the strong
coupling constant $\alpha_s$ in perturbative QCD
\begin{equation}
  \label{eq:5}
  C_n^{theo} = C_n^{(0)} + \frac{\alpha_s}{4\pi} C_n^{(1)} + \left
    (\frac{\alpha_s}{4\pi} \right ) ^2 C_n^{(2)} + \left
    (\frac{\alpha_s}{4\pi} \right ) ^3 C_n^{(3)} + \cdots \, .
\end{equation}
To perform the  calculation the initially present propagator-type diagrams
are expanded in $q^2/4 M^2$ where $q$ is the external photon momentum and $M$
the mass of the heavy quark. This expansion leads to tadpole diagrams
  to which the techniques presented in the previous Section can be
  applied.

The Taylor coefficients have been calculated at ${\cal O}(\alpha_s^2)$,
i.e. three-loop order, in  \cite{Chetyrkin:1995ii,Chetyrkin:1997mb,Boughezal:2006uu,Maier:2007yn}. At four loops the first moment has been
calculated in  \cite{Chetyrkin:2006xg,Boughezal:2006px,Sturm:2008eb} and the  second
moment in \cite{Maier:2008he}, see \cite{Grozin:2004ez,Czakon:2007qi}
for higher moments for purely fermionic contributions. Using the second
moment the analysis presented in \cite{Kuhn:2007vp}, see also \cite{Kuhn:2001dm,Boughezal:2006px}, can be upgraded to include the now
complete knowledge of the second moment. Using this newly available
information the error of the bottom quark mass can be reduced to obtain
\begin{equation}
  \label{eq:6}
    m_b(10 \mbox{GeV}) = 3.607(19) \mbox{GeV}   \, ,
\end{equation}
which has to be compared with old value of 
\begin{equation}
  \label{eq:7}
    m_b(10  \mbox{GeV}) = 3.609(25)  \mbox{GeV} . 
\end{equation}
In the case of the charm quark one finds 
\begin{equation}
  \label{eq:8}
  m_c(3 \mbox{GeV}) = 0.976(16)\mbox{GeV}
\end{equation}
instead of 
\begin{equation}
  \label{eq:9}
  m_c(3 \mbox{GeV}) = 0.979(22)\mbox{GeV} \, .
\end{equation}
Even though the shift in the absolute value is quite small the error gets
reduced by around 25\%. 

Since the different moment weights the experimental input differently
precise knowledge of the second moment allows for a consistency check of
the method used for the extraction of the quark masses. Especially in
the case of the bottom quark the second moment is favored over the
first one since it is more sensitive to the threshold region and the
narrow resonances.

\section{CONCLUSION}
\label{sec:conclusion}
We presented a short review of the current status of an integral part of
multi-loop calculations, namely the reduction to master integrals. Using
recent Groebner basis inspired methods in combination with Laporta-like
approaches most reduction problems can be solved in an efficient way and
the most problematic part is again the calculation of the master
integrals. As an example we presented the calculation of the second
moment of the heavy quark current correlator at four-loop order and its
application to the extraction of the charm- and bottom-quark masses from
experimental data.

\begin{acknowledgments}
This work was supported by the DFG through SFB/TR~9. P.M. thanks A.~Maier,
P.~Maierh\"ofer and A.V. Smirnov for collaboration and reading the manuscript.
\end{acknowledgments}

\end{document}